\documentclass[letterpaper]{article} 
\usepackage{aaai25}  
\usepackage{times}  
\usepackage{helvet}  
\usepackage{courier}  
\usepackage[hyphens]{url}  
\usepackage{graphicx} 
\usepackage{natbib}  
\usepackage{caption} 
\usepackage{algorithm}
\usepackage{algorithmic}
\usepackage{booktabs}
\usepackage{multirow}
\usepackage{xcolor}
\usepackage{array}              
\definecolor{customblue}{HTML}{466A9E}   
\definecolor{customgreen}{HTML}{38855B}
\newcolumntype{M}[1]{>{\raggedright\arraybackslash}m{#1}}
\newcolumntype{C}[1]{>{\raggedright\arraybackslash}m{#1}}

\usepackage{newfloat}
\usepackage{listings}
\DeclareCaptionStyle{ruled}{labelfont=normalfont,labelsep=colon,strut=off} 
\floatstyle{ruled}
\newfloat{listing}{tb}{lst}{}
\floatname{listing}{Listing}
\newcommand{\ail}{$AI^2L$}

\title{Human-in-the-loop or AI-in-the-loop? \\ Automate or Collaborate?}
\author{
   Sriraam Natarajan\textsuperscript{\rm 1}, Saurabh Mathur\textsuperscript{\rm 1}\equalcontrib,
    Sahil Sidheekh\textsuperscript{\rm 1}\equalcontrib, \\ Wolfgang Stammer\textsuperscript{\rm 2,}\textsuperscript{\rm 3}, Kristian Kersting\textsuperscript{\rm 2,}\textsuperscript{\rm 3}
}
\affiliations{
    \textsuperscript{\rm 1} University of Texas at Dallas, \textsuperscript{\rm 2} Technical University of Darmstadt,\\
    \textsuperscript{\rm 3} Hessian Center for Artificial Intelligence (hessian.ai), Darmstadt, Germany

}

\usepackage{bibentry}

\begin{document}

\maketitle

\begin{abstract}
Human-in-the-loop (HIL) systems have emerged as a promising approach for combining the strengths of data-driven machine learning models with the contextual understanding of human experts. However, a deeper look into several of these systems reveals that calling them HIL would be a misnomer, as they are quite the opposite, 
namely AI-in-the-loop (\ail{}) systems: 
the human is in control of the system, while the AI is there to 
support the human. We argue that existing evaluation methods often overemphasize the machine (learning) component's performance, neglecting the human expert's critical role. Consequently, we propose an \ail{} perspective, which recognizes that the human expert is an active participant in the system, significantly influencing its overall performance. By adopting an \ail{} approach, we can develop more comprehensive systems that faithfully model the intricate interplay between the human and machine components, leading to more effective and robust AI systems.
\end{abstract}

%

\section{Introduction}

Since the time of the ``advice taker"~\cite{McCarthy1959,McCarthy68programswith}, there has been a significant interest in building human-allied AI systems. Different paradigms and different techniques such as active learning~\cite{settlestr09}, knowledge-based learning~\cite{kbann,kbsvm}, explanatory interactive learning~\cite{SchramowskiSTBH20,Stammer21rightconcept}, advice-taking~\cite{dejongMooney86,BaffesMooney96,Odom15AAAI}, weak/distant supervision~\cite{natarajan2015effectively, ratner2017snorkel}, human feedback~\cite{maclin05,wiewiora03,ngetal99,griffith2013policy}
and preference elicitation~\cite{boutilier2002pomdp, boutilier2004cp,ChenP12,toni2024personalized} etc., to name a few, have been developed for this important and challenging task. Many of these directions have been presented under the umbrella of {\em Human-in-the-loop} (HIL) systems. We take a deeper look at these systems and ask the question if they are truly HIL systems.

To understand the difference, let us consider two simple examples: 
\begin{enumerate}
\item An AI system that recommends content to users (say videos). If a human intervenes in such a system, they provide feedback/guidance by either correcting inappropriate content (as a trusted ally) or by providing malicious/inappropriate content (as an adversary). In either case, the AI agent optimizes its internal function, considers the human feedback, and decides on the appropriate action (in this case, showing the relevant content). 
\item As a second example, consider an AI system that assists a physician who is treating a diabetic patient for a knee injury and prescribes oral steroids to mitigate pain. AI could now intervene based on its internal objective function, domain constraints, and knowledge and suggest that since the patient has diabetes, the physician should reconsider their recommendation. The physician can then inform the system that the patient is in acute pain and reducing that is important or, in contrast, that the patient did not inform the physician and hence will change her prescription.
\end{enumerate} 
Indeed, in either of these cases, the AI system interacts with the human expert, assimilates knowledge, updates its constraints, makes internal computations, and then provides suggestions. However, although these two systems appear quite similar at the outset, there is a crucial difference in the role of decision-making authority and control. In the former case, AI is in charge of decision-making and takes additional inputs from the human expert. Arguably, these can be ``richer'' inputs than treating the human as a ``mere labeler''. Still, the human is not the decision-maker, while the AI actually is. In the latter case, the human is in control of the full system. The presence of AI inside this system only makes the process more efficient and possibly more effective. However, the full system exists independent of the presence of AI. This difference is critical. 

Consider the problem of evaluation. Clearly, evaluating a system based only on its performance (say accuracy or some other function of precision/recall) will benefit the HIL system but not the \ail{} system, as argued by~\citet{van2024algorithms}. More importantly, during deployment, the issues that the \ail{} systems must address and reason with can be significantly different from HIL systems. 

In the rest of this blue sky paper, we first present these two systems in greater detail showing their similarities and differences. We argue strongly that when designing systems that operate in the presence of human experts, the designers of these systems must clearly understand the conditions under which these systems operate and then decide whether a HIL or \ail{} system is appropriate for the task at hand. After all, there is no one ring to rule them all!

\begin{figure*}
    \centering
    \includegraphics[width=0.95\linewidth]{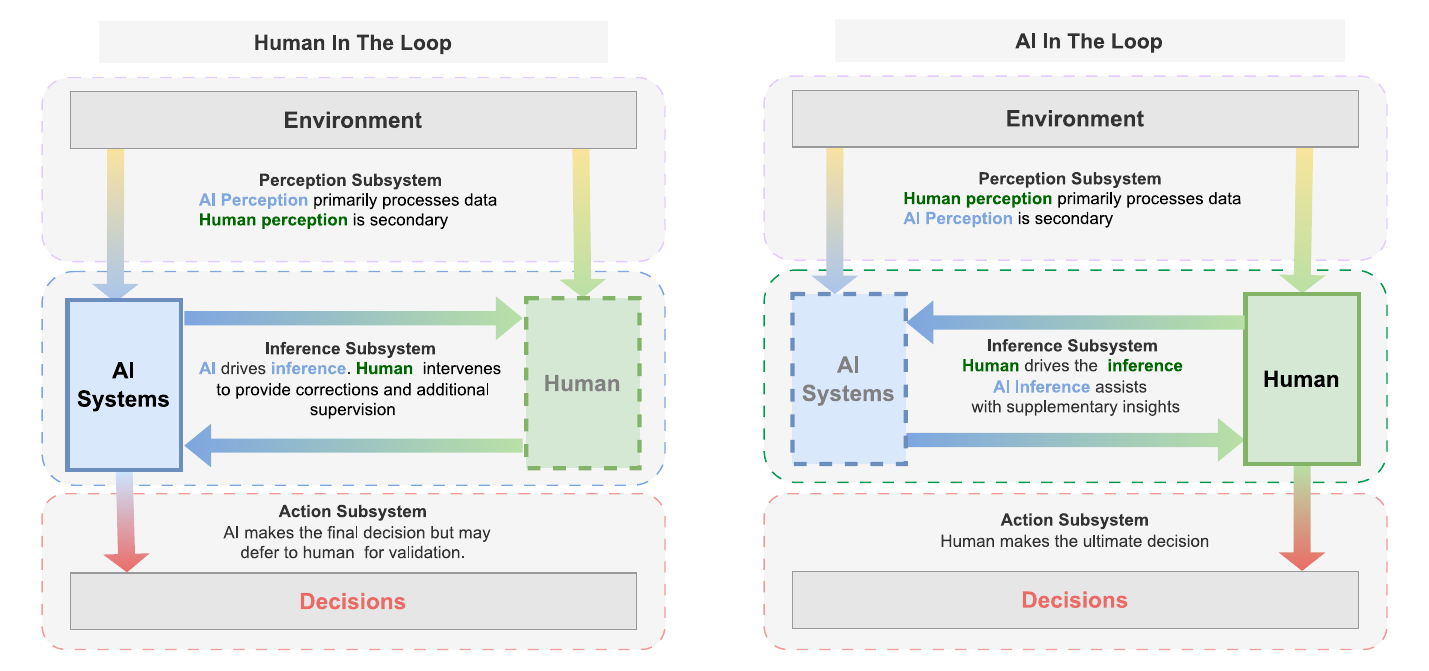}
    \caption{A comparison of human in the loop (left) and AI in the loop (right) systems. In human-in-the-loop systems, AI systems drive the inference and decision-making process, but humans intervene to provide corrections and supervision. In AI-in-the-loop systems, humans make the ultimate decisions, while AI systems assist with perception, inference, and action. }\label{fig:ail-vs-hil}
\end{figure*}

\section{Human-in-the-loop}

The Human-in-the-loop paradigm for developing AI systems typically treats humans either as data-labeling oracles~\cite{settlestr09} or as a source of domain knowledge~\cite{mosqueira2023human}. Humans primarily function as oracles in the Active Learning (AL, see~\citet{settlestr09}) paradigm or as weak supervisors~\cite{ratner2017snorkel}, providing labels for unlabeled data instances the model finds uncertain, ambiguous, or missing. These approaches are valuable for domains with large amounts of unlabeled data and for which annotation is costly or time-consuming. While the system controls the learning process by selecting which instances are presented to the human for labeling, AL aims to improve model accuracy with fewer training examples. However, this approach relies on the assumption that humans prefer acting as efficient labeling machines, falling short of making \textit{human use of human beings}~\cite{wiener1988human}.

Machine teaching (MT, see 
\citet{simard2017machineteachingnewparadigm}) focuses on making the teachers who build machine learning models more effective, rather than improving just the learning algorithms. Over the past two decades, however, most research has centered on developing powerful (deep) learning algorithms for handling abundant data~\cite{nature2015deep}. However, as machine learning expands to address more varied and often rather short-term tasks, the scarcity and cost of skilled teachers have become limiting factors. Inspired by the evolution of programming in the 1980s and 1990s, machine teaching emphasizes principles like problem decomposition, modularity, and process discipline. It draws parallels with programming, highlighting version control, semantic data exploration, and the expressiveness of teaching languages as key aspects of effective machine teaching.

Specifically, the MT paradigm posits humans as teachers who guide machine learning models to acquire specific knowledge. This allows domain experts to create effective models in the absence of large data sets without deep ML expertise. Such knowledge-intensive learning methods have a long history in AI, from John McCarthy's work in the 1960s~\cite{McCarthy68programswith} to explanation-based learning~\cite{dejongMooney86}, theory refinement~\cite{BaffesMooney96}, and inductive logic programming and relational learning \cite{muggletonR94,deraedt08}. The key motivation is that experts have extensive knowledge in their respective fields, which many data-driven ML techniques~\cite{mitchell97,svmBook,boostingBook} do not fully exploit. 

In fact, the use of advice in various forms has produced successful algorithms, particularly in reinforcement learning~\cite{maclin05,wiewiora03,ngetal99}, where advice is used as reward shaping. In supervised learning, advice is typically provided as feature selection or inductive bias on initial models. In graphical models, advice is used as an initial structure that is refined~\cite{heckerman1998tutorial}. On the other hand, knowledge-based neural networks~\cite{kbann} and support vector machines~\cite{kbsvm,kunapuli2010online}, inverse RL~\cite{kunapuli2013guiding}, relational models~\cite{Odom15AAAI,odom2018human,odom2016active} and probabilistic model learning~\cite{AltendorfEtAl2005, Campos2008QIBN, yang2013knowledge, KiGBKokel2020, Mathur2023KICN, Mathur2024Credal} have explored combining knowledge and data to handle systematic noise. While adaptation specifics may differ, all of these methods can take advice as Horn clauses, convert them to their corresponding representation, and learn by using them as constraints.

The common denominator in all of these systems is that the AI module is in control of the decision process and the human inputs are essentially used to ``guide" the model to a better (possibly local) optimum. This scenario is explored in the LHS of Figure~\ref{fig:ail-vs-hil}. The performance of HIL systems is typically measured from the system's perspective and is based on accuracy, precision, recall, or a function of these metrics. Issues of trust (due to inherent biases), and credibility are important challenges in these systems. These mainly stem from the human expert's biases and data biases and include, but are not limited to confirmation bias, conformity bias, attribution bias, affinity bias, halo effect, cognitive bias, and racial and gender bias to name a few. A common issue of HIL systems is thus the danger of manipulation by an adversary that provides incorrect advice, ultimately requiring the system to model the credibility of the human experts to make effective decisions.

\begin{table*}[ht!]
\centering

\caption{
Examples of subtasks across various domains and their classification as \textbf{\textcolor{customblue}{Automate (HIL)}} or \textbf{\textcolor{customgreen}{Collaborate ($AI^2L$)}}.
}
\label{tab:examples}
\resizebox{.9\linewidth}{!}{
\begin{tabular}{@{}C{3cm} C{3.5cm} C{3cm} C{7cm}@{}}
\toprule
\textbf{Domain} & \textbf{Task} & \textbf{\shortstack{\textbf{\textcolor{customblue}{Automate (HIL)}} or \\ \textbf{\textcolor{customgreen}{Collaborate ($AI^2L$)}?}}} & \textbf{Description} \\ \midrule

\multirow{2}{*}{Medicine} & 
Early diagnosis of Alzheimer's disease & 
\textcolor{customblue}{Automate} & 
\textcolor{customblue}{AI analyzes patient data (e.g., MRI) and detects anomalies.} \\ \cmidrule(l){2-4} 
& 
Treatment plan formulation & 
\textcolor{customgreen}{Collaborate} & 
\textcolor{customgreen}{Physician selects and tailors the final treatment plan from a set of candidates generated by AI.} \\ \specialrule{1pt}{0pt}{0pt}

\multirow{2}{*}{Automobile} & 
Route planning from source to destination & 
\textcolor{customblue}{Automate} & 
\textcolor{customblue}{AI computes optimal routes using real-time traffic and weather data, requiring minimal human input.} \\ \cmidrule(l){2-4} 
& 
Driving in high-density, urban environment & 
\textcolor{customgreen}{Collaborate} & 
\textcolor{customgreen}{Human drivers navigate complex traffic, assisted by AI for tasks like collision avoidance, lane change and adaptive cruise control.} \\ \specialrule{1pt}{0pt}{0pt}

\multirow{2}{*}{Logistics} & 
Shipping cost forecasting based on historical data & 
\textcolor{customblue}{Automate} & 
\textcolor{customblue}{AI uses prior data to predict shipping costs; human manager can correct previous mistakes, provide additional context and advice.} \\ \cmidrule(l){2-4} 
& 
Inventory management & 
\textcolor{customgreen}{Collaborate} & 
\textcolor{customgreen}{Human decides on procurement strategy based on AI's estimates of stock/reorder points.} \\ \specialrule{1pt}{0pt}{0pt}

\multirow{2}{*}{Manufacturing} & 
Detecting known quality issues in product & 
\textcolor{customblue}{Automate} & 
\textcolor{customblue}{AI automatically detects product defects; human inspectors confirm corrections if needed.} \\ \cmidrule(l){2-4} 
& 
Quality assurance and compliance & 
\textcolor{customgreen}{Collaborate} & 
\textcolor{customgreen}{Humans design quality assurance strategies based on quality issue patterns identified by AI.} \\ \specialrule{1pt}{0pt}{0pt}

\multirow{2}{*}{Finance} & 
Fraud detection & 
\textcolor{customblue}{Automate} & 
\textcolor{customblue}{AI analyzes the data and flags suspicious activities/transactions; humans confirm.} \\ \cmidrule(l){2-4} 
& 
Investment advisory & 
\textcolor{customgreen}{Collaborate} & 
\textcolor{customgreen}{Human advisor decides on the final strategy using AI-provided market analyses and recommendations.} \\ \specialrule{1pt}{0pt}{0pt}

\multirow{2}{*}{Education} & 
Automated grading of assignments & 
\textcolor{customblue}{Automate} & 
\textcolor{customblue}{AI gives feedback on assignments based on commonly seen mistakes (e.g., software bugs); human instructor provides advice, sets grading schema and policies, and reviews uncertain cases.} \\ \cmidrule(l){2-4} 
& 
Curriculum Updates & 
\textcolor{customgreen}{Collaborate} & 
\textcolor{customgreen}{Teachers update and adapt lessons to student needs based on AI's analyses of student learning trajectories across time.} \\ \bottomrule
\end{tabular}
}
\end{table*}

\section{AI-in-the-loop}
Many applications of AI and machine learning involve interactions with humans. Humans may provide input to a learning algorithm, including input in the form of labels, demonstrations, corrections, rankings, or evaluations. They could give such input while observing the algorithm’s outputs, potentially in the form of feedback, predictions, or demonstrations. However, although humans are arguably an integral part of the learning process, traditional machine learning systems are agnostic to the fact that inputs/outputs are from/for humans. In fact, machine learning is often conceived --- in particular in applications of other scientific disciplines such as medicine --- in a very impersonal way, with algorithms working autonomously on passively collected data.

In contrast, interactive machine learning (IML, \citet{fails2003interactive, amershi2014power, michael2020interactive,WARE2001281,WinNT,Teso23frontiers}) represents a shift toward greater human involvement and shared control in the learning process. Humans can assume various roles in IML, from domain experts and data scientists to non-expert users. This flexibility allows for a more dynamic interplay between humans and machines, assigning tasks based on individual strengths and capabilities. Unlike the algorithm-centric focus of AL, IML systems require a human-centered approach to evaluation that incorporates additional judgments such as calibration, fairness, and explainability alongside traditional performance metrics that merely measure the system's conformity to past data. Beyond their direct involvement in training, humans are also the ultimate users of AI systems. This requires evaluating AI systems not only for their functionality but also for their usability and usefulness to human users.

In keeping with the paradigm shift of IML systems, we argue for a change in thinking about these systems. Specifically, consider the RHS of Figure~\ref{fig:ail-vs-hil}. It can be easily seen that while the system is very similar to the HIL system, there are crucial differences. While in HIL systems AI is in control, in \ail{}, the human is at the center of the system and fully in control. 
Despite these critical differences, \textit{e.g.}, in the deployment of safe AI-in-the-loop systems, we observe that many of the existing literature simply consider these two systems to belong to one group. However, we strongly argue for their separation and will highlight their differences next.

First, consider the issues of reliability and trust in these systems. The biases in \ail{} systems are mainly due to algorithmic and model biases that are reflective of the data bias. Moreover, trust issues in these systems are drastically different from those of HIL systems. While the credibility assessment of the human teacher is crucial in HIL, transparency of the system, its explainability, and interpretability are crucial in \ail{} systems~\cite{rossHD17,tesoK19,lipton2018mythos, rudin2019stop, SchramowskiSTBH20,stammer2024neural}. Moreover, trust is much more nuanced in \ail{} systems than HIL systems because the human user is in control and is unlikely to trust a system if it does not align with their expectations. Hence, the user's confirmation bias could potentially be reflected in their willingness to trust the underlying \ail{} system. From the perspective of credibility in the context of \ail{}, instead of assigning credibility to humans (as in HIL systems), typically \ail{} systems compute the credibility of data sources, for example, the different modalities or knowledge bases from which the data are being extracted. 

Second, the evaluations of these systems are human-centric and are mostly aligned with the broader goals of the environment in which they operate. Although metrics such as Precision, Recall, Accuracy, and F-scores etc.~are still relevant in understanding the performance of the AI system, arguably much more emphasis should be placed on the impact on the human who is at the center of the system. Hence, ablation studies are more important in such systems to evaluate the impact of the different components on the overall system, for instance, on specific health outcomes. While in HIL systems, ablation studies are useful, in \ail{} systems, they are essential. Typically, some other important aspects of the evaluation of these \ail systems are the interpretability, explainability\cite{sreedharan2022explainable}, interactive capabilities~\cite{zahedi2023}, and generalizability of these systems~\cite{wust2024pix2code}.

Above all, the most important consideration concerns the system itself. Is the AI system necessary in this task that is already performed by the human? If so, what is the potential impact of the AI system, efficiency or efficacy, or both? What are the potential hazards of using an AI system in this task? How can the improvements in the systems be measured objectively? Furthermore, are the biases due to the model or the data? How credible are the data sources that helped create the \ail{} system? These questions must be answered deliberately before the system can be deployed. 

This is the crux of our argument -- {\em instead of considering every system in which a human is present to be a HIL system, it is imperative to understand the type of system, their evaluation criteria, and potential implications of its deployment in greater detail.}

In fact, \ail{} systems are related to the vision of bridging explainable and advisable AI and achieving a human-AI symbiosis~\cite{zahedi2021,sreedharan2022explainable,kambhampatiSVZG22,zahedi2023,kambhampati2020}. 
Both emphasize that the human and the computer are both in the loop, and AI becomes a co-adaptive process, in which a human is changing AI behavior, but the human also adapts to use AI more effectively and adapts their data and goals in response to what is learned using machine learning. \ail{} systems, however, emphasize the need to move beyond the train/test evaluation paradigm of static, non-contextualised benchmarks,
toward user- or even population-specific metrics and evaluation protocols close to the real-life requirements
of society.  

\section{Discussion}
HIL and AI2L systems differ in three key aspects, namely, control, source of bias, and evaluation. The first difference is that HIL systems are generally autonomous AI agents that might seek specific help from humans, while AI2L systems constitute an intervention in a human decision-making process. The AI component in AI2L presents the human with a summary of information synthesized from multiple sources, a set of possible allowable actions, and their possible consequences (e.g., a human selecting a single decision from a set of MPE solutions). This difference in control configurations results in differing sources of bias. While HIL systems are primarily vulnerable to bias in historical data and domain knowledge used in model construction, AI2L systems are also vulnerable to biases arising from human interpretation of the AI's output. Finally, while HIL systems are evaluated using AI-centered metrics such as accuracy, precision, and recall, AI2L systems require a more holistic approach to evaluation, taking into account the human-AI interaction, the overall goals of the decision-making process, and considerations such as fairness that cannot be fully quantified. 
Table \ref{tab:examples} presents a few concrete examples from diverse domains, grounding the distinctions between HIL and $AI^2L$ systems in practical, real-world contexts. It categorizes sample subtasks across areas such as Medicine, Automobile, Logistics, Manufacturing, Finance, and Education, illustrating whether they are best addressed through automation with HIL oversight or through collaborative engagement via $AI^2L$ systems.

The choice of HIL or \ail{} perspective influences the aspects of a system that are abstracted away during design and evaluation. Designing systems as HIL when they should be understood as \ail{} can result in abstraction errors ~\cite{selbst2019fairness}, allow for modeling the conscious and unconscious biases that arise due to humans, evaluate the system on incorrect or inappropriate criteria, or have serious consequences after deployment. For instance, using an HIL system to regulate an exceedingly complex stochastic system like the human body~\cite{BeerCybernetics67} might overlook crucial contextual details necessary for clinical decision-making. In contrast, the \ail{} perspective recasts system deployment as an intervention in existing processes, allowing evaluation strategies to be more closely aligned with end goals such as improving health outcomes~\cite{van2024algorithms} and minimizing harmful social outcomes~\cite{mohla2021material}.

In summary, the appropriate problem domains for the HIL and \ail{} systems are typically not separated but nested. Automation is most effective in well-defined contexts, while human intervention is most needed in not yet defined or undefinable contexts. Hence, zooming in on a domain would result in a HIL problem and zooming out would give us an \ail{} problem, \textit{e.g.,} identifying drug-drug interactions is a reasonably well understood context, making it appropriate for HIL while general medical diagnosis is not as clearly defined, making it more appropriate for \ail{}. Additionally, while effective software engineering requires active human decision-making~\cite{johnson2024ai}, automating some well-defined sub-problems such as static analysis and vulnerability detection~\cite{yadavally2023partial} can help reduce the software engineer's cognitive load and lead to better quality software.

While our discussions are motivated from the perspective of supervised learning, the frameworks are agnostic to the type of learning performed. Our arguments for the difference in the two systems apply directly to unsupervised learning, reinforcement learning, planning, continual learning, and meta-learning to name a few. While specific adaptations differ, the idea of humans in the center or AI in the center applies broadly across these different settings, 
see e.g.~\cite{Delfosse24scobots}.
Or consider current foundation models trained in a self-supervised fashion. While \ail{} systems focus on AI-supported human agency, current foundation models lend themselves more easily to HIL settings where AI is the primary actor, and humans intervene to monitor or enhance results. Their ability to generalize across tasks and perform with minimal additional training allows humans to take on roles of oversight and feedback rather than constant, direct involvement---the user still gives a thumbs up or down on the text generated by a large language model. Although they may pick up information about how to collaborate with humans, without an understanding of the user’s goals, beliefs, or uncertainties, even foundation models are likely to remain reactive rather than truly collaborative partners. By improving their ``theory of mind'', however, foundation models could offer more contextual, meaningful suggestions that integrate with human workflows and align better with social values and the evolving needs of human users.

In short, moving from HIL to \ail{} is likely to help build AI systems where AI truly enhances human expertise, resulting in smarter, more resilient solutions that thrive on collaboration, not automation. Doing so, however, requires the AI community to rethink its evaluation methodology.

\section{Acknowledgements}
SN sincerely thanks Rao Kambhampati for his insightful discussions that led to a deeper dive into the differences between HIL and \ail{} systems. SN, SM, and SS gratefully acknowledge the generous support by AFOSR award FA9550-23-1-0239, the ARO award W911NF2010224 and the DARPA Assured Neuro Symbolic
Learning and Reasoning (ANSR) award HR001122S0039. KK and WS  acknowledge that they benefited from the ``ML2MT'' project (Volkswagen Stiftung) and the 
3AI project from the Hessian Ministry of Science and Arts (HMWK). KK and WS also benefited from the European Project ``Tango'' (Grant Agreement no. 101120763); the views and opinions expressed are, however, those of the author(s) only and do not necessarily reflect those of the European Union or the European Health and Digital Executive Agency (HaDEA); neither the European Union nor the granting authority can be held responsible for them.

\bibliography{aaai25}

\end{document}